# Audio Scene Classification with Deep Recurrent Neural Networks

*Huy Phan*⋆†, *Philipp Koch*⋆, *Fabrice Katzberg*⋆, *Marco Maass*⋆, *Radoslaw Mazur*⋆
*and Alfred Mertins*⋆

⋆Institute for Signal Processing, University of Lübeck
†Graduate School for Computing in Medicine and Life Sciences, University of Lübeck
{phan,koch,katzberg,maass,mazur,mertins}@isip.uni-luebeck.de

## Abstract

We introduce in this work an efficient approach for audio scene classification using deep recurrent neural networks. An audio scene is firstly transformed into a sequence of high-level label tree embedding feature vectors. The vector sequence is then divided into multiple subsequences on which a deep GRU-based recurrent neural network is trained for sequence-to-label classification. The global predicted label for the entire sequence is finally obtained via aggregation of subsequence classification outputs. We will show that our approach obtains an F1-score of 97.7% on the LITIS Rouen dataset, which is the largest dataset publicly available for the task. Compared to the best previously reported result on the dataset, our approach is able to reduce the relative classification error by 35.3%.

## 1. Introduction

The ability to recognize a surrounding environment using acoustic signals has potential for many applications. Therefore, the challenge of audio scene classification (ASC) recently gained great attention from the research community [1, 2].

There is an ongoing methodology trend in dealing with the task, which is shifting from conventional classification techniques to modern deep learning methods. This trend can be seen in the recent DCASE 2016 challenge [2]. State-of-the-art performances on different benchmark datasets have been reported by several works which pursue deep neural networks (DNNs) [3] and convolutional neural networks (CNNs) [4, 5, 6]. However, despite their top performance, these network variants are not capable of modeling sequences. Therefore, it is arguable that there is room for improvement beyond DNNs and CNNs by explicitly modeling the sequential dynamics of acoustic scene signals. However, this is very challenging due to complex sound composites of acoustic scenes. It is evidenced by that recurrent neural networks (RNNs), e.g. Long Short-Term Memory (LSTM) [7], which are highly capable of sequence modeling, have been shown to be inferior to CNN competitors [8, 9]. They are often used in combination with CNNs to benefit from their feature learning ability [10]. To our knowledge, there is no prior work succeeding in training standalone RNNs with on par or better performance than those of CNNs in the same ASC benchmarks.

We propose an approach that successfully trains deep RNNs for ASC and leads to the state-of-the-art performance on the LITIS Rouen dataset [11]. Audio scenes are of complex content which typically consists of background noise mixed with rich foreground sounds. In general, both background noise and foreground sounds can be used to characterize a scene. However, foreground sounds usually occur in an arbitrary order, making hidden sequential patterns hard to uncover. In the proposed approach, the complex audio scenes are firstly transformed and reduced into meta-class likelihoods via a label tree embedding (LTE) to expose their sequential patterns. As a result, a scene instance is transformed into a sequence of LTE feature vectors on which training RNNs is efficient. Due to the arbitrary occurrences of foreground sounds, we divide the whole vector sequences into smaller subsequences to enhance the repeatability of patterns. The network training and evaluation finally take place on a subsequence level, followed by an aggregation step using some voting schemes. We also investigate different factors that may influence the classification performance, namely the number of network layers, subsequence length, overlapping degree, different LTE feature types, different aggregation methods, multi-stream setting of RNNs, and classification calibration with a linear SVM. The proposed approach bears resemblances to previous works which used LTE features for ASC [12, 13, 4]. However, while these works did not treat the ASC task as a sequential modeling problem, explicitly modeling of LTE feature vector sequences is the key to obtain the state-of-the-art performance in the proposed approach.

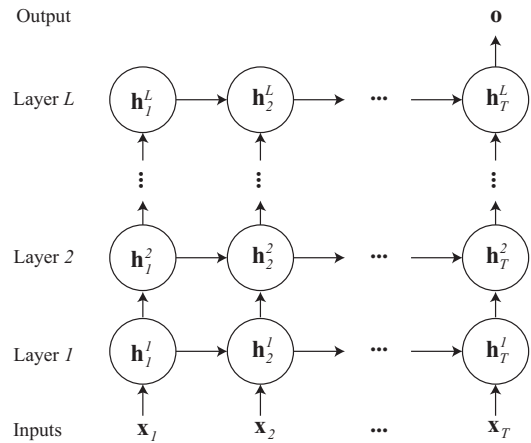

Figure 1: *Illustration of the proposed deep GRU-based RNN architecture.*

## 2. Deep RNN for sequence classification

### 2.1. Deep GRU-based RNN architecture

The proposed deep RNN is presented in this section for sequence-to-label classification. Given a sequence $\mathbf{X} = (\mathbf{x}_1, \ldots, \mathbf{x}_T)$ of $T$ feature vectors $\mathbf{x}$, a standard RNN iterates over the individual input feature vectors and computes the sequence of hidden state vectors $\mathbf{H} = (\mathbf{h}_1, \ldots, \mathbf{h}_T)$. At a

time step $t$, where $1 \leq t \leq T$, $\mathbf{h}_t$ is computed as

$$\mathbf{h}_t = \mathcal{H}(\mathbf{x}_t, \mathbf{h}_{t-1}), \quad (1)$$

where $\mathcal{H}$ denotes the hidden layer function. In our proposed network, we employ the Gated Recurrent Unit (GRU) cell [14] in which the function $\mathcal{H}$ is implemented by the compound of following functions:

$$\mathbf{r}_t = \mathrm{sigm}\left(\mathbf{W}_{xr}\mathbf{x}_t + \mathbf{W}_{hr}\mathbf{h}_{t-1} + \mathbf{b}_r\right), \quad (2)$$

$$\mathbf{z}_t = \mathrm{sigm}\left(\mathbf{W}_{xz}\mathbf{x}_t + \mathbf{W}_{hz}\mathbf{h}_{t-1} + \mathbf{b}_z\right), \quad (3)$$

$$\tilde{\mathbf{h}}_t = \tanh\left(\mathbf{W}_{xh}\mathbf{x}_t + \mathbf{W}_{hh}\left(\mathbf{r}_t \odot \mathbf{h}_{t-1}\right) + \mathbf{b}_h\right), \quad (4)$$

$$\mathbf{h}_t = \mathbf{z}_t \odot \mathbf{h}_{t-1} + (1 - \mathbf{z}_t) \odot \tilde{\mathbf{h}}_t. \quad (5)$$

In above equations, the $\mathbf{W}$ variables denote the weight matrices and the $\mathbf{b}$ variables are the biases. The $\mathbf{r}$, $\mathbf{z}$, and $\tilde{\mathbf{h}}$ variables represents the reset gate vector, the update gate vector, and the new hidden state vector candidate, respectively. The $\odot$ operator denotes the element-wise vector product. Since our classification problem is sequence-to-label, the network output is determined via the final state vector $\mathbf{h}_T$:

$$\mathbf{o} = \mathbf{W}_{hy}\mathbf{h}_T + \mathbf{b}_y. \quad (6)$$

Although the GRU architecture has been demonstrated better performance compared to the LSTM one [7] in many sequence modeling tasks [15, 16], the adoption of GRU architecture in this work is mainly due to its lower computational cost. It is not necessary to be optimal for ASC.

Similar to [17], in order to construct a deep RNN, we stack multiple RNN hidden layers on top of each other as demonstrated in Fig. 1. Assuming that the deep RNN has $L$ layers in total, the hidden state sequence of a lower layer is treated as the input sequence for the upper one:

$$\mathbf{h}_t^\ell = \mathcal{H}\left(\mathbf{h}_t^{\ell-1}, \mathbf{h}_{t-1}^\ell\right), \quad (7)$$

where $1 \leq \ell \leq L$. In particular, $\mathbf{H}^0 = \mathbf{X}$ for the first layer.

The output of the deep RNN is then given by

$$\mathbf{o} = \mathbf{W}_{hy}\mathbf{h}_T^L + \mathbf{b}_y. \quad (8)$$

The network output is subsequently presented to a softmax layer to compute the predicted probability $\hat{\mathbf{y}}$ over the class labels.

### 2.2. Network training

The deep RNN is trained using $N$ training examples $\{(\mathbf{X}_1, \mathbf{y}_1), \ldots, (\mathbf{X}_N, \mathbf{y}_N)\}$ where $\mathbf{y}$ denotes a binary one-hot encoding vector. The training procedure is accomplished by minimizing the cross-entropy error over the training examples:

$$E(\boldsymbol{\theta}) = -\frac{1}{N}\sum_{n=1}^{N}\mathbf{y}_n \log\left(\hat{\mathbf{y}}_n(\mathbf{X}_n, \boldsymbol{\theta})\right) + \frac{\lambda}{2}\|\boldsymbol{\theta}\|_2^2. \quad (9)$$

In (9), $\boldsymbol{\theta}$ denotes the network trainable parameters and the hyper-parameter $\lambda$ is used to compromise the error term and the $\ell_2$-norm regularization term. We also exploit dropout [18] on the output of the last layer for further regularization. That is, the elements of the output vector are randomly set to zeros with a pre-defined probability. The optimization is performed using the *Adam* optimizer [19].

### 2.3. Calibration with support vector machine (SVM)

For classification, SVM usually leads to better generalization in comparison with the standard softmax, thanks to its maximum margin property [20]. SVMs has been used in combination with CNNs for classification [21, 22, 4]. Similarly, after training the deep RNN, we calibrate the final classifier by employing a linear SVM in replacement for the softmax layer.

The output vectors of the network extracted for the training examples are used to train the linear SVM classifier which is subsequently applied to classify those output vectors extracted for the test examples. By doing this, we have treated the deep RNN as a feature extractor which was trained to produce good representations for the classification task at hand. Finally, the raw SVM scores can be calibrated and converted into a proper posterior probability as in [23] when it is needed.

## 3. Deep RNNs for ASC

### 3.1. Feature extraction

Although deep RNNs can be trained on low-level features, such as for speech phoneme recognition [17], it is arguably inefficient to do so for the ASC task. The main reason is that the content of acoustic scenes is far richer and much less structured in comparison to those of speech phonemes. Alternatively, we use high-level LTE features for training as in [13]. Using the LTE features, we have transformed and reduced the rich audio content of the scenes into meta-class likelihoods [12] to expose their sequential information. As a result, the RNNs can be trained more easily.

Given a target scene dataset consists of $C$ classes, we represent each 30-second audio snippet as a sequence of LTE feature vectors. An audio snippet is first decomposed into small segments of length 250 ms with a hop size of 125 ms to obtain 238 segments. A low-level feature vector, such as MFCCs, is then extracted for each segment. Afterwards, we transform the per-segment low-level feature vector into an LTE feature vector as in [12]. More specifically, the low-level feature vectors are used to construct a binary label tree which indexes $(C-1) \times 2$ meta-classes at left and right child nodes of the tree. A low-level feature vector $\mathbf{x}$ is then mapped into an LTE feature vector $\Psi(\mathbf{x}) \in \mathbb{R}^{(C-1) \times 2}$ given by

$$\Psi(\mathbf{x}) = \left(\psi_1^l(\mathbf{x}), \psi_1^r(\mathbf{x}), \ldots, \psi_{C-1}^l(\mathbf{x}), \psi_{C-1}^r(\mathbf{x})\right), \quad (10)$$

where $\psi_i^l(\mathbf{x})$ and $\psi_i^r(\mathbf{x})$ denote the posterior probabilities of it belonging to two meta-classes at the left and right child nodes of the split node index $i$, respectively.

As in [13], we investigate different low-level feature sets for LTE feature learning. The first set consists of 64 Gammatone cepstral coefficients [24] extracted in the frequency range of 20 Hz to half of the sampling frequency. For the second set, we extract 60 MFCC coefficients as in [25]. The third set consists of 20 log-frequency filter bank coefficients, their first and second derivatives, zero-crossing rate, short-time energy, four sub-band energies, spectral centroid, and spectral bandwidth [26]. For feature extraction, a 250-ms segment is further divided into frames of 50 ms long with a hop size of 25 ms. The feature extraction is performed on the frame level. In turn, the feature vector for the whole 250-ms segment is calculated by averaging the per-frame feature vectors.

Furthermore, for each low-level feature set, we extract two LTE features corresponding to the presence/absence of background noise and concatenate them on segment-wise basis as they can complement each other [4, 13]. The background noise is subtracted using the minimum statistics estimation and subtraction method [27] when necessary. As a result, three LTE feature sequences, namely *LTE-Gam*, *LTE-MFCC*, and *LTE-Log*, are obtained for a scene snippet.

### 3.2. ASC with single-stream RNNs

Given three LTE channels (i.e. *LTE-Gam*, *LTE-MFCC*, and *LTE-Log*), we train deep RNNs on individual ones, namely *RNN-Gam*, *RNN-MFCC*, and *RNN-Log*, respectively. Furthermore, we exploit training a deep RNN, namely *RNN-Fusion*, on the combination of multiple LTE channels. Feature fusion allows the network to leverage patterns across different LTE channels. For this purpose, the LTE feature vectors of different channels are simply concatenated at every time step.

The sequences of 238 time steps are divided into 8 subsequences of 32 time steps (equivalent to 4 seconds) without overlapping, except for the last subsequence. The training and classification are accomplished on the subsequences. The influence of the subsequence length and the overlapping degree will also be studied in the experiments. The classification label for an entire sequence consisting of $M$ subsequences ($M = 8$ in this case) are determined by majority voting (MV) on the predicted labels of the constituent subsequences. In addition, we investigate three probabilistic voting schemes: maximum (Max. PV), additive (Add. PV), and multiplicative (Mul. PV). Let $\mathbf{P}^m = (P_1^m, \ldots, P_C^m)$ denote the classification probabilities obtained for a subsequence index $m$. The classification likelihood $\mathbf{P} = (P_1, \ldots, P_C)$ obtained by Max. PV, Add. PV, and Mul. PV schemes are given by

$$P_i = \max(P_i^m) \quad \text{for} \quad 1 \le m \le M \text{ and } 1 \le i \le C, \quad (11)$$

$$P_i = \frac{1}{M} \sum_{m=1}^{M} P_i^m \quad \text{for} \quad 1 \le i \le C, \quad (12)$$

$$P_i = \frac{1}{M} \prod_{m=1}^{M} P_i^m \quad \text{for} \quad 1 \le i \le C, \quad (13)$$

respectively. The predicted label $\hat{c}$ is then determined by

$$\hat{c} = \underset{i}{\operatorname{argmax}} P_i \quad \text{for} \quad 1 \le i \le C. \quad (14)$$

### 3.3. ASC with multi-stream RNNs

We also investigate fusion of different networks trained on individual LTE feature channels (i.e. *RNN-Gam*, *RNN-MFCC*, and *RNN-Log*) in a multi-stream setting, namely *RNN-Multi*. The idea is inspired by multi-stream networks that have been shown their efficiency in different classification tasks [28, 29, 4]. Again, the fusion is accomplished using MV, Max. PV, Add. PV, and Mul. PV strategies. Given $K = 3$ individual RNN streams, Max. PV, Add. PV, and Mul. PV schemes can be re-written as

$$P_i = \max(P_i^{m,k}) \quad \text{for} \quad 1 \le m \le M \text{ and } 1 \le i \le C, \quad (15)$$

$$P_i = \frac{1}{MK} \sum_{m=1}^{M} \sum_{k=1}^{K} P_i^{m,k} \quad \text{for} \quad 1 \le i \le C, \quad (16)$$

$$P_i = \frac{1}{MK} \prod_{m=1}^{M} \prod_{k=1}^{K} P_i^{m,k} \quad \text{for} \quad 1 \le i \le C, \quad (17)$$

respectively, where $k \in \{RNN\text{-}Gam, RNN\text{-}MFCC, RNN\text{-}Log\}$. The final predicted label is determined as in (14).

Table 1: *Parameters of the deep RNN architecture.*

| Parameter | Value |
|---|---|
| The number of layers $L$ | varied in $\{1, 2, 3, 4\}$ |
| Size of hidden state vector | 256 |
| Learning rate for Adam optimizer | $10^{-4}$ |
| Dropout rate | 0.1 |
| Regularization parameter $\lambda$ | $10^{-3}$ |

## 4. Experiments

### 4.1. LITIS-Rouen dataset

The experimental dataset consists of 3026 examples of 19 scene categories [11]. Each class is specific to a location such as a train station or an open market. The audio recordings have a duration of 30 seconds and a sampling rate of 22050 Hz. The dataset has a total duration of 1500 minutes. To our knowledge, this is the largest available dataset for the ASC task. We follow the training/testing splits in the seminal work [11] and report average performances over 20 splits.

### 4.2. Parameters

The parameters needed for LTE feature extraction are similar as those in [12]. The parameters associated with the proposed RNN architecture are given in Table 1. We varied the number of layers to investigate its influence. The RNNs were trained with 100 epochs and a batch size of 100. Finally, the hyper-parameter $C$ of the SVMs used for calibration was fixed to 1.0.

### 4.3. Experimental results

#### 4.3.1. Performance of different deep RNNs

The performances in terms of F1-score obtained by different RNNs with $L = 2$ layers (the best case) are shown in Fig. 2. Overall, among different aggregation methods, Mul. PV appears to be the best one. Compared to others, this voting scheme strongly favors and suppresses classification likelihoods of categories that have consistent and diverged subsequence classification results, respectively [30].

It can also be seen from Fig. 2 that good performances can be obtained with the individual LTE feature types (i.e. *RNN-Gam*, *RNN-MFCC*, and *RNN-Log*). For instance, *RNN-Gam* achieves a highest F1-score of 96.6% which is already better than that obtained with the best CNN with multiple LTE types reported in [13] (i.e. 96.5%). This result implies that appropriate sequential modeling is important for ASC. The performance is further improved by integrating multiple LTE types either with a simple feature concatenation or the multi-stream setting. More specifically, absolute gains of 1.1% and 0.7% are obtained by *RNN-Fusion* and *RNN-Multi* over *RNN-Gam*, respectively. These results also confirm the similar findings in [4, 12].

While the standard softmax is convenient for network training, calibrating the final classification step with linear SVMs yields significant performance improvements as shown in Fig. 2. The absolute gains (averaged over different aggregation methods) of 0.8%, 1.5%, 0.8%, 0.8%, and 0.4% are obtained by the SVM-calibrated *RNN-Gam*, *RNN-MFCC*, *RNN-Log*, *RNN-Fusion*, and *RNN-Multi* compared to uncalibrated counterparts, respectively.

#### 4.3.2. Influence of the number of layers

We show in Table 2 the classification performances obtained by the RNNs which have their numbers of layers varying from 1

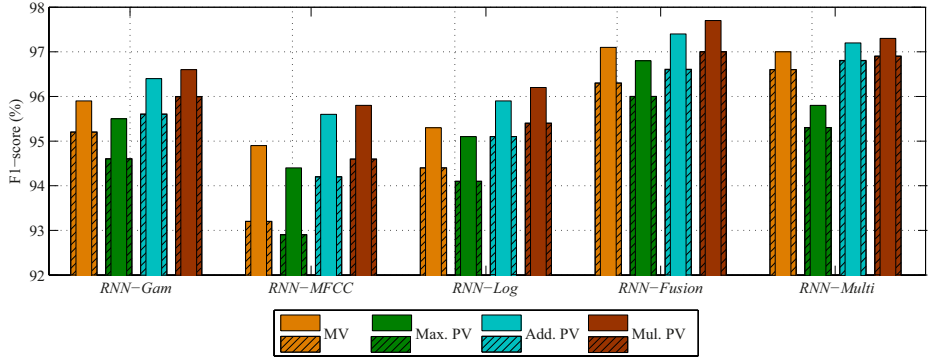

Figure 2: *F1-scores (%) obtained by 2-layer RNNs with different aggregation methods. The solid/stripe-pattern bars correspond to the results with/without SVM calibration.*

to 4. Note that these results are obtained with SVM calibration and the Mul. PV scheme. As can be seen from Table 2, there are only small variations in the F1-scores. Nevertheless, $L = 2$ is likely the most appropriate and the RNNs appear to be deteriorated with larger numbers of layers.

*4.3.3. Influence of the subsequence length and overlap*

The subsequence length and the overlapping degree may also have their own effects. To examine them, we reduced the subsequence length by 50% (i.e. from 32 to 16 time steps) and repeated the experiment with the best setting for *RNN-Fusion* (e.g. with Mul. PV and SVM calibration). This network yields an F1-score of 97.3% which is inferior to the same RNN with 32 time steps with an F1-score loss of 0.2% absolute. In another experiment, we kept the subsequence length unchanged and used an overlap of 50%. As a result, the F1-score is dropped by 0.4% absolute, reducing from 97.7% to 97.3%.

Negative effects introduced by the overlap is likely due to the fact that they cause the GRU cells overwrite their memory more often. Furthermore, the subsequence length should be considered as a trade-off parameter. Short subsequences will force a network to learn detailed patterns. In contrast, generalized patterns may be hard to be found with too long subsequences. Although not shown here, we experimentally experienced with significant performance drops when using the whole sequences as inputs.

*4.3.4. Performance comparison*

To illustrate the efficiency of the proposed approach, we show in Table 3 the performances of our RNNs in comparison to those reported in previous works. Our RNNs are with 2 layers, Mul. PV aggregation, and SVM calibration. The performances are reported in average class-wise precision, F1-score, and overall accuracy for a proper comparison.

As can be seen from Table 3, our *RNN-Gam*, *RNN-Multi*, and *RNN-Fusion* systems consistently outperform all competitors over all evaluation metrics. Furthermore, *RNN-Fusion* sets the state-of-the-art performance and outperforms the best results in prior works (i.e. CNN-Fusion6 [4]) by 1.4%, 1.2%, and 1.2% absolute on precision, F1-score, and accuracy, respectively. This performance gain is equivalent to a relative classification error reduction of 35.3%.

## 5. Conclusions

In summary, we proposed an approach using deep GRU-based RNNs for audio scene classification. The audio scenes were presented as sequences of label tree embedding features. The sequences were subsequently decomposed into multiple subsequences on which the RNNs were trained for sequence-to-label classification. The classification for a long sequence was finally obtained by aggregating classification outputs of its constituent subsequences. We achieved the state-of-the-art performance on the LITIS Rouen dataset with the proposed approach. Furthermore, we also demonstrated the influence of different related factors that affects the network training.

Table 2: *F1-score (%) obtained by RNNs with different numbers of layers L.*

|  | RNN-Gam | RNN-MFCC | RNN-Log | RNN-Fusion | RNN-Multi |
|---|---|---|---|---|---|
| 1 layer | 96.7 | 95.8 | 96.1 | 97.6 | 97.2 |
| 2 layers | 96.6 | 95.8 | 96.2 | 97.7 | 97.3 |
| 3 layers | 96.5 | 95.5 | 96.2 | 97.5 | 97.3 |
| 4 layers | 96.6 | 95.5 | 96.0 | 97.3 | 97.2 |

Table 3: *Performance comparison on the LITIS Rouen dataset.*

| System | Prec. | F1-score | Acc. |
|---|---|---|---|
| *RNN-Fusion* | **97.5** | **97.7** | **97.8** |
| *RNN-Multi* | **97.1** | **97.3** | **97.4** |
| *RNN-Gam* | **96.4** | **96.6** | **96.7** |
| *RNN-MFCC* | 95.4 | 95.8 | 96.0 |
| *RNN-Log* | 95.9 | 96.2 | 96.4 |
| CNN-Fusion6 [4] | *96.3* | *96.5* | *96.6* |
| LTE1-Fusion3 [4] | 95.5 | 95.7 | 95.8 |
| Scene-LTE + Speech-LTE [12] | 95.9 | 96.2 | 96.4 |
| HOG [11] | 91.7 | – | – |
| DNN+MFCC [31] | 92.2 | – | – |
| HOG+SPD [32] | 93.3 | 92.8 | 93.4 |
| Sparse NMF [33] | – | 94.1 | – |
| Convolutive NMF [33] | – | 94.5 | – |
| Kernel PCA [33] | – | 95.6 | – |
| FisherHOG+ProbSVM [29] | – | – | 96.0 |

## 6. Acknowledgements

This work was supported by the Graduate School for Computing in Medicine and Life Sciences funded by Germanys Excellence Initiative [DFG GSC 235/1].